\documentclass[prl,twocolumn,a4paper,groupaddress,showpacs]{revtex4}
\usepackage{graphicx}

\begin{document}

%\draft

\title{Mermin-Ho vortex in ferromagnetic spinor Bose-Einstein condensates}

\author{T. Mizushima, K. Machida and T. Kita$^{1}$}
\affiliation{Department of Physics, Okayama University,
         Okayama 700-8530, Japan, \\
         $^{1}$Division of Physics, Hokkaido University,
         Sapporo 060-0810, Japan}
\date{\today}
%\maketitle

\begin{abstract}
The Mermin-Ho and Anderson-Toulouse
coreless non-singular vortices are demonstrated to be thermodynamically stable
in ferromagnetic spinor Bose-Einstein condensates with the hyperfine state $F=1$.
The phase diagram is established in a plane of the rotation drive vs the total
magnetization by comparing the energies for other competing non-axis-symmetric
or singular vortices. Their stability is also checked by evaluating collective modes. 
\end{abstract}


\maketitle

%\narrowtext

%%%%%%%%%%%%%%%%%%%%%%%%%%%%%%%%%%%%%%%%%%
%%%%%%%%%%%%%
Topological structure plays an important and decisive 
role in various research fields, ranging from condensed 
matter physics to high energy physics. They provide  a 
common framework to connect diverse
fields, enhancing mutual understanding\cite{jayaraman}.

Recent advance of experimental techniques on Bose-Einstein condensation
(BEC) prompts us to closely and seriously look into theoretical 
possibilities which were mere imagination for theoreticians in this field.
This is particularly true for spinor BEC where all hyperfine states of  
an atom Bose-condensed simultaneously, keeping these ``spin'' states
degenerate and active. Recently, Barrett at al \cite{barrett} have succeeded in cooling 
$^{87}$Rb with the hyperfine state $F=1$ by all optical methods without
resorting to a usual magnetic trap in which the internal degrees of freedom
is frozen. Since the spin interaction of the  $^{87}$Rb atomic system is 
ferromagnetic, based on the refined calculation of the atomic interaction parameters
by Klausen at al\cite{klausen}, we now obtain concrete examples of the three 
component spinor
BEC ($F=1$, $m_F=1,0,-1$) for both antiferromagnetic 
($^{23}$Na)\cite{stenger} and ferromagnetic interaction cases.
In the present spinor BEC the degenerate internal degrees of freedom play an
essential role to determine the fundamental physical properties. There is 
a rich variety of topological defect structures, which are already predicted in 
the earlier studies\cite{ohmi,ho} on the spinor BEC.
These are followed by 
others\cite{yip,leonhardt,stoof,marzlin,anglin,kurihara,martikainen,isoshima1,isoshima2}
 who examine these topological structures more
closely, such as skyrmion, monopole, meron or axis-symmetric or non axis-symmetric
vortices both for antiferromagnetic and ferromagnetic cases.

Superfluid $^{3}$He is analogous to the spinor BEC where 
the neutral Cooper pair possesses
the orbital and spin degrees of freedom, thus the order parameter
is a multi-component\cite{salomaa}. Among various topological structures,
the Mermin-Ho (MH)\cite{mermin} and 
Anderson-Toulouse (AT)\cite{anderson} vortices
of a coreless and non-singular $l$-vector texture are proposed in  $^{3}$He-A phase. 
These are  an extremely interesting object to study if they exist. 
The MH vortex is expected to spontaneously appear in a cylindrical vessel 
without any external 
rotation as an equilibrium state because the rigid wall boundary imposes the 
$l$-vector perpendicular to the vessel wall.
The MH vortex is stable also under slow rotation because of
their non-singular coreless structures\cite{he3}.

 A similar topological structure, called skyrmion in general is proposed 
 in the spinor BEC. Khawaja and Stoof\cite{stoof} study a  skyrmion in the 
 ferromagnetic BEC with $F=1$ and conclude that the skyrmion 
 is not a thermodynamically stable object without rotation. Once created,
 its radius shrinks to vanish in spite of ingenious 
 proposals\cite{marzlin,anglin,kurihara,martikainen} as
 to how to created it and to detect it.

 Here we demonstrate that in the ferromagnetic spinor BEC with  $F=1$
 trapped in a harmonic potential the Mermin-Ho vortex and
 Anderson-Toulouse vortex are thermodynamically stable for 
 a certain region of the external rotation frequency and the 
magnetization of the system. These vortices are stabler than 
 the singular vortices and other possible non-axis symmetric vortices.
 This comes about because as a general tendency the ferromagnetic
 interaction leads to spatial phase separation of the three components 
 $(\phi_{1}, \phi_{0}, \phi_{-1})$. Then by assigning different
 winding numbers $\langle 0,1,2\rangle$   to these components in this order,
 they can be arranged so as to be effectively phase-separated in 
 the radial direction, resulting in a concentric layered structure 
 under a given magnetization. The central region is occupied by $\phi_{1}$,
  $\phi_{0}$ is  in the intermediate region, and $\phi_{-1}$
 is in the outer layer. This ingenious non-singular coreless vortex 
 acquires the angular momentum to effectively lower the total 
 energy under rotation.
 
 We start with the standard Hamiltonian by Ohmi and Machida\cite{ohmi}, and
 Ho\cite{ho}: $
%   H =
%   \int\! d{\bf r} [
%      \sum_{ij}\Psi_i^{\dagger}
%          \left\{h({\bf r}) - \mu_{i}\right\}
%      \Psi_j \delta_{ij}
%      + \frac{g_{\rm n}}{2} \sum_{ij}
%         \Psi_i^{\dagger} \Psi_j^{\dagger} \Psi_j \Psi_i          
%      + \frac{g_{\rm s}}{2} \sum_{\alpha}
%          \sum_{ijkl} \Psi_i^{\dagger}\Psi_j^{\dagger}
%          (F_{\alpha})_{ik}(F_{\alpha})_{jl}
%          \Psi_k \Psi_l
%      - {\bf \Omega} \cdot \sum_{j} \Psi_j^{\dagger}({\bf r} \times {\bf p})\Psi_j
%      ]
%-----
H_{rot} = H - \int\! d{\bf r} {\bf \Omega} \cdot 
              \sum_{j} \Psi_j^{\dagger}({\bf r} \times {\bf p})\Psi_j
$, 
\begin{eqnarray*}
H = \int\! d{\bf r} [
                      \sum_{ij}\Psi_i^{\dagger}
                      \left\{h({\bf r}) - \mu_{i}\right\}
                      \Psi_j \delta_{ij}
                      + \frac{g_{\rm n}}{2} \sum_{ij}
                        \Psi_i^{\dagger} \Psi_j^{\dagger} \Psi_j \Psi_i \nonumber \\
                      + \frac{g_{\rm s}}{2} \sum_{\alpha}
                        \sum_{ijkl} \Psi_i^{\dagger}\Psi_j^{\dagger}
                             (F_{\alpha})_{ik}(F_{\alpha})_{jl}
                        \Psi_k \Psi_l
                    ]
\end{eqnarray*}
where 
$
   h({\bf r}) = - \frac{\hbar^2 \nabla^2}{2m} + V({\bf r})$, $
   g_{\rm n} = \frac{4 \pi \hbar^2}{m} \cdot \frac{a_0 + 2a_2}{3}$,
 $g_{\rm s} = \frac{4 \pi \hbar^2}{m} \cdot \frac{a_2 - a_0}{3} $.
The subscripts are $\alpha = (x,y,z)$ and $i,j,k,l = (0, \pm1)$
corresponding to the above three species.
The chemical potentials for the three components;
$\mu_i$ $(i = 0, \pm1)$ 
satisfy $\mu_1 + \mu_{-1} = 2 \mu_{0}$.
The scalar field $V({\bf r})=\frac{1}{2} m (2\pi\nu_r)^2 (x^2 + y^2)$ 
is the external confinement potential
such as an optical potential.
The scattering lengths $a_{0}$ and $a_{2}$ characterize collisions between
atoms with the total spin 0 and 2 respectively. As mentioned, the 
recent refined estimate\cite{klausen} for $^{87}$Rb concludes it ferromagnetic ($a_0>a_2$)
and $g_s/g_n=-0.01\sim -0.005$. The external rotation frequency $\Omega$
is normalized by the harmonic confining frequency.
It readily leads to the Gross-Pitaevskii (GP) equation
extended to the three component
order parameters:
$
   [
      \left\{
          h({\bf r}) - \mu_i + g_{\rm n} \sum_k |\phi_k|^2 
      \right\} \delta_{ij} 
      + g_{\rm s} \sum_{\alpha}\sum_{kl} \{
         (F_{\alpha})_{ij} (F_{\alpha})_{kl} \phi_k^{\ast} \phi_l
      \}
   ] \phi_j =
    0.
$
% %%%%%%%%%%%%%%%%%%%%%%%%% %%%%%%%%%
%
%begin{eqnarray}
 %  [
%      \left\{
%          h({\bf r}) - \mu_i + g_{\rm n} \sum_k |\phi_k|^2 
%      \right\} \delta_{ij} 
%&&\nn\\
 %     + g_{\rm s} \sum_{\alpha}\sum_{kl} \{
 %        (F_{\alpha})_{ij} (F_{\alpha})_{kl} \phi_k^{\ast} \phi_l
%      \}
 %  ] \phi_j &=& 0.
%\label{eq:gp}
%\end{eqnarray}
% %%%%%%%%%%%%%%%%%%%%%%%%%%%%%%%%
These coupled equations for the $j$-th condensate wave function
$\phi_j=\langle \Psi_j \rangle$  $(j = 0, \pm1)$ are used to calculate
various properties of a vortex in the following.

It is important to realize that in a real experimental situation 
the total number  $N=\int d{\bf r} \Sigma_j|\phi_j|^2$ and the 
total magnetization  $M=\int d{\bf r} \Sigma_jj|\phi_j|^2$ are fixed. 
In our case these quantities are fixed by adjusting the chemical 
potential ($\mu_0$) and the fictitious magnetic field ($\mu'=\mu_1-\mu_0$).
The extended GP equation is solved numerically with two
different methods for a two-dimensional disk: One is not to assume axis-symmetry.
This calculation is backed up by the computation where the axis-symmetry
is assumed. The actual calculations are done by discretizing two-dimensional
space into 51$\times$51 mesh. The vortex configuration is characterized by 
the combination of the winding number of each component. We denote it as 
$\langle w_1,w_0,w_{-1}\rangle$   where integers $w_1,w_0,w_{-1}$  are the winding 
numbers of $\phi_1,\phi_0,\phi_{-1}$ respectively where $w$ means the phase
change by $2\pi w$ when the wave function goes around the phase singularity.

We have performed extensive search to find a stable vortex, starting with possible 
vortex configurations, covering a wide range of the ferromagnetic interaction
strength $g_s/g_n=0\sim -0.02$ and examining various axis-symmetric and 
non-axis-symmetric vortices (see the classification of possible vortices in 
the axis-symmetric case\cite{isoshima1,isoshima2}).
We use the following parameters: 
the mass of a $^{87}$Rb atom m=1.44$\times 10^{-25}$kg, 
the trapping frequency $\nu_r$=200Hz, 
and the area density $n_z$=2.0$\times 10^3 / \mu$m.
The results displayed here are $g_s/g_n=-0.02$.

The Mermin-Ho (MH) vortex is described by 
$(\phi_1,\phi_0,\phi_{-1})=\sqrt{n(r)}
(\cos^2{\beta\over 2}, \sqrt 2e^{i\phi}\sin{\beta\over 2}\cos{\beta\over 2},
e^{2i\phi}\sin^2{\beta\over 2})$
where the bending angle $\beta(r)$ is $0\leq\beta(r)
\leq\pi$, $\phi$ is the polar angle in polar coordinates. The spin direction is denoted by
the so-called $l$-vector\cite{ohmi} given by $\mbox{\boldmath $l$}(r)={\hat z}\cos\beta+
\sin\beta (\cos\phi{\hat x}+\sin\phi{\hat y})$ where $\beta$ varies from
$\beta(0)=0$ to $\beta(R)=\frac{\pi}{2}$ for MH and to $\beta(R)=\pi$ for AT ($R$ is 
the outer boundary of the cloud). Thus the spin moment is flared out to the radial
direction and at the circumference it points outward for MH and downwards for AT
(see for schematic $l$-vector structure Fig.18 in Ref.\cite{salomaa} ).
These vortices have the winding number combination $\langle 0,1,2\rangle$  in our notation.

%===================== Figure 1 ======================
\begin{figure}
  \begin{tabular}{cc}
    \includegraphics[width=5cm]{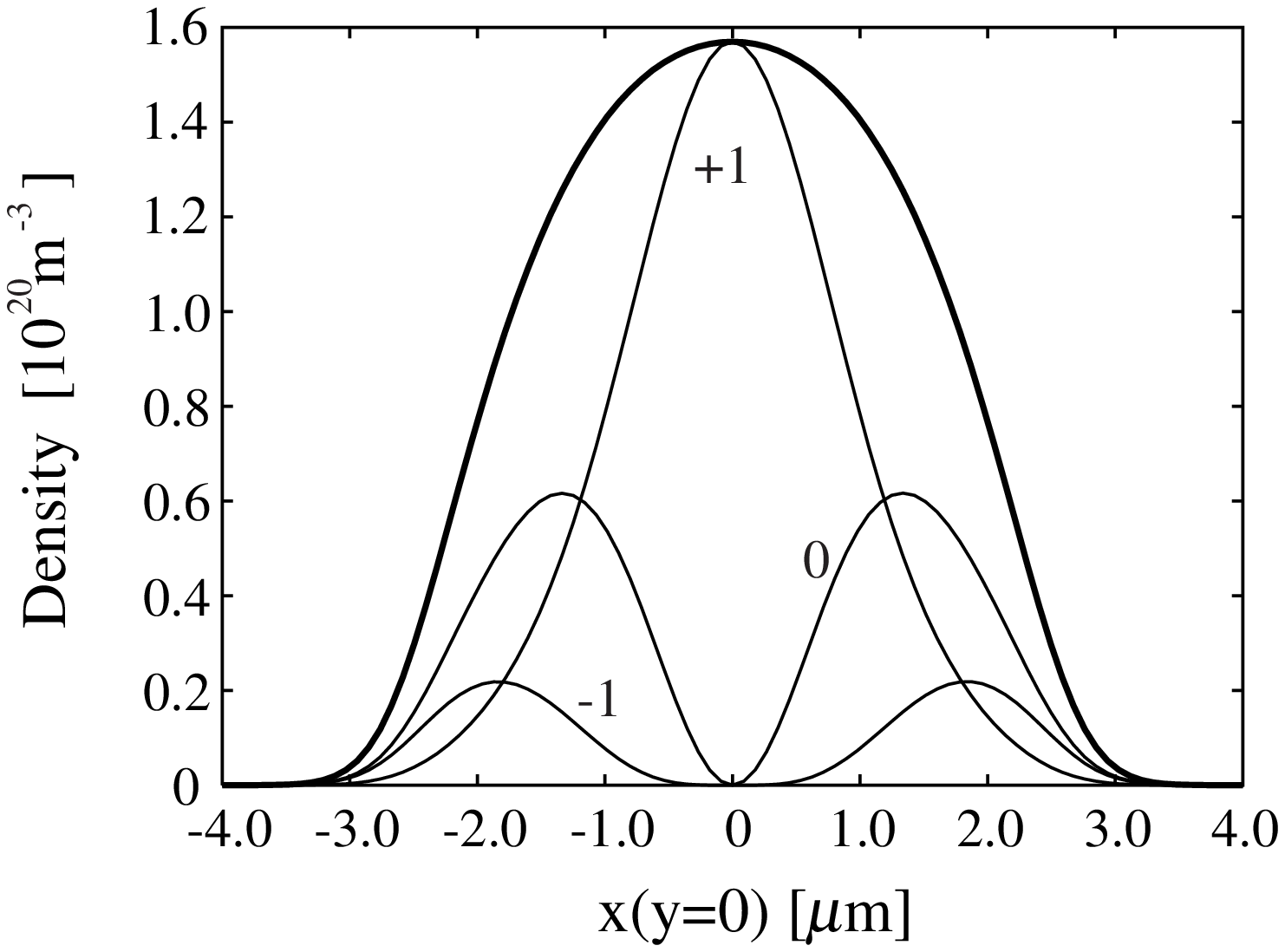}
    &
    \includegraphics[width=3.5cm]{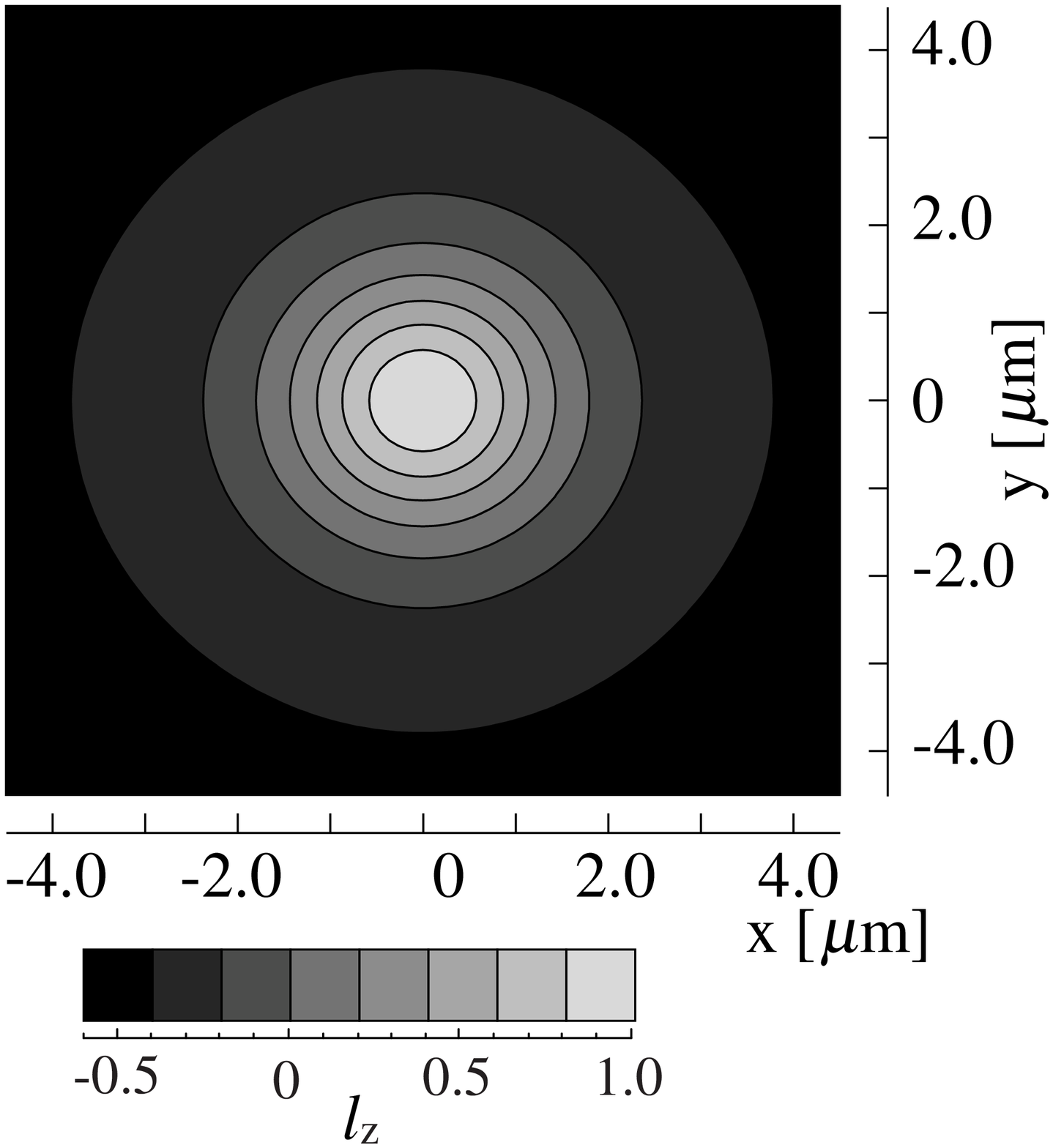}\\
    (a)
    & 
    (b) 
  \end{tabular}
\caption{\label{fig:012dns}
Density profile of the condensates (a) 
and the density map of $l_z$-vector (b) 
for the $\langle 0,1,2\rangle$ vortex at $\Omega=0.35$
and M/N=0.21.
The bold line is the total density $\sum_j | \phi_j |^2$ 
and the thin lines show the density of each component $|\phi_j|^2$.
}
\end{figure}
%===================== Figure 1 ======================/

 We show the calculated MH vortex in Fig.1 where the axis-symmetric
 density profiles for each
 component and the density map of $l_z$ are displayed. It is
 seen that the central region of the harmonic trap is occupied by $\phi_1$
 with zero winding number $w_1=0$. The  $\phi_0$ component which has a
 singularity with $w_0=1$ at $r=0$ is pushed outward while the $\phi_{-1}$ component
 occupies the circumference region because of $w_{2}=2$. Note that 
 in the small $r$ region the singular vortex with $w$ behaves like  $\propto r^w$.
 The resulting total density is coreless and non-singular. It has a smooth density variation
 described by a Gaussisn form except for the outermost region.

 In Fig.2 we show the spatial dependence of the $l_z$-component 
 along the radial direction, namely, the spatial dependence of the bending
 angle $\beta(r)$ for the MH vortex. As the magnetization $M$ decreases,
 the local magnetization changes from positive to negative values through zero.
 It means that the $l$-vector in the vortex flares out radially to orient almost horizontally
 $\beta(r=R)={\pi\over 2}$ for $M/N \sim 0.5$ and to point down  for 
 $\beta(r=R)=\pi$ for $M/N \sim 0$. The former (latter) corresponds literally to 
 the Mermin-Ho (Anderson-Toulouse) vortex. This is simply because 
 as $M$ decreases, the spin-down component $\phi_{-1}$ with $w=2$ increases in
 the outer region. Thus we can control these MH and AT vortices by merely
 changing the total magnetization. We notice that in superfluid $^3$He-A phase
 the stability of the MH and AT vortices is guaranteed by the boundary 
 condition at the wall of a vessel where the $l$-vector is constraint to be 
 parallel to the wall surface\cite{salomaa}. Here the situation is completely different;
 we impose no constraint on the $l$-vector direction. These vortex configurations
 are created naturally under the condition of a given total number and magnetization,
 both of which are well controlled in a harmonic trap experiment.
 
 This difference in the two systems is rather remarkable and interesting. 
 In our case MH or  AT vortex comes about purely due to the spin interaction, which 
 is written as $\propto |g_s||2\phi_1(r)\phi_{-1}(r)-\phi_0(r)|^2$.
 This term favors the mutual phase segregation\cite{isoshima1,isoshima2}.
 If $\phi_1(r)\neq0$ and 
$\phi_{-1}(r)\neq0$, $\phi_0(r)$ comes in, explaining the concentric layered structure.
 In fact, in the antiferromagnetic case the similar calculation shows that the MH and AT vortices 
 are never stabilized under a similar condition\cite{isoshima2,mizushima}.
 
 We have done extensive search for determining the region for the $\langle 0,1,2\rangle$  
 vortex of the 
 MH and AT in the plane of $\Omega$ vs $M/N$.
 We check our computation by starting with a certain initial vortex configuration,
 allowing non-axis-symmetric vortex. We end up with the axis-symmetric
 MH and AT vortices where these are stable. The stability for these vortices
 is also examined from the two aspects: one is the global stability and the other is
 local stability in the energy landscape. The global stability means comparing the
 relative energy of various vortices and vortex free state to select out the lowest energy state.
 The local stability is discussed shortly.

%===================== Figure 2 ======================
\begin{figure}
%\begin{center}
\includegraphics[width=5.5cm]{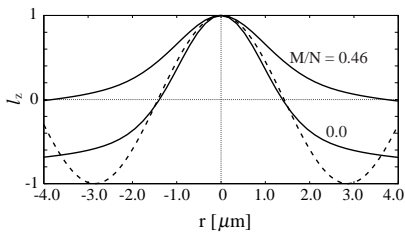}
%\end{center}
\caption{\label{fig:lz_M}
Spatial dependence of the $l_z$-component along the radial direction 
at M/N=0, 0.46, and $\Omega=0.37$.
The dashed line shows cos$\beta(r)$ with the bending angle
$\beta(r)=\pi r/R$ ($R=2.85\mu m$)
for the AT vortex.
}
\end{figure}
%===================== Figure 2 ======================/ 
 
 The resulting phase diagram is displayed in Fig.3 where a large 
 area is occupied by the $\langle 0,1,2\rangle$  vortex, including MH and AT.
 Near $M\sim0$ the non-axis-symmetric  $\langle 1,1,1\rangle$  vortex  and 
 near $M/N\sim 1$, $\langle 1,0,-1\rangle$  vortex are stabilized. We find a large
 empty region in the intermediate $M/N$ region where no vortex and 
 vortex-free state are stabilized at all because the phase separation in
 the ferromagnetic case prevents forming a uniform mixture 
of the three components even when the circulation is absent for the
vortex-free state. This is in contrast with the antiferromagnetic case
in which everywhere is occupied by a stable phase.

It is noted that the focused frequency region $\Omega=0.38$ corresponds
to the single vortex region in the scalar BEC case\cite{isoshima3} beyond which multiple
vortex state must be seriously considered, thus the present single vortex
consideration is justified in the lower frequencies.

%===================== Figure 3 ======================
\begin{figure}
%\begin{center}
\includegraphics[width=8cm]{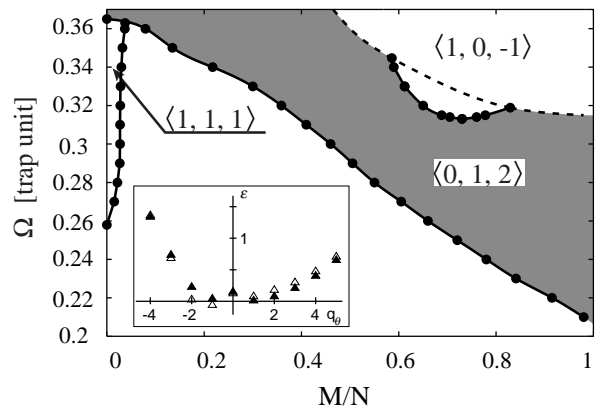}
%\end{center}
\caption{
   \label{fig:Fdiagram}
Phase diagram in the ferromagnetic state.
The lowest eigenvalues along $q_{\theta}$ at $\Omega$=0.3 
are displayed in the inset. 
Triangles: M/N=0.2 and filled triangles: 0.75.
}
\end{figure}
%===================== Figure 3 ======================/

We examine various competing vortices, including a singular vs non-singular
and axis-symmetric vs non-axis-symmetric vortices.
In particular, since for axis-symmetric vortices the winding number 
combination of the three components $\langle w_1,w_0,w_{-1}\rangle$  is restricted to
$2w_0=w_1+w_{-1}$, we exhaust all the possible vortices with  $w_1,w_0,w_{-1}$
smaller than unity\cite{isoshima2}. Note that MH and AT belong to this axis-symmetry category.

The $\langle 1,1,1\rangle$  vortex shown in Fig. 4 is non-singular and non-axis-symmetric
which is stable in low $M/N$.
This $\langle 1,1,1\rangle$  vortex is advantageous because (A) it does not contain the 
higher winding number. It generally leads to collapse fewer winding number vortex,
such as $2\rightarrow1+1$ vortices. (B) The overall condensation energy is gained 
by placing their singularities off the trap center where the potential energy
is minimum. (C) As $\Omega$ increases, the two separate singularities of $\phi_1$
and $\phi_{-1}$ adjust their mutual distance from the center so as to maximize
the angular momentum ${\vec L}$ in order to save the energy 
$-{\vec \Omega}\cdot{\vec L}$. In this sense this configuration is flexible against
varying $\Omega$. These reasons explain why this vortex survives along the 
$\Omega$-axis near $M/N\sim0$ in Fig.3.

In contrast, the $\langle 0,1,2\rangle$  vortex contains the higher winding number 2
for  $\phi_{-1}$, which makes it less advantageous against the $\langle 1,1,1\rangle$  vortex.
However, upon varying $M$, this vortex is quite flexible by
adjusting the particle numbers for the three components; As $M$ increases,
the number of $\phi_1$ grows smoothly relative 
to the rests. The non-winding $\phi_1$
component works as a ``pinning'' center for the remaining $\phi_0$
and  $\phi_{-1}$. In particular,  $\phi_{-1}$ with $w_{-1}=2$ is stabilized by
the presence of $\phi_1$. Therefore as $M$ increases or $\phi_1$ grows, the
MH and AT vortices become more and more stable. This explains that the 
critical frequency, or the lower phase boundary of MH in Fig.3 becomes less
as $M$ increases. As for the $\langle 1,1,1\rangle$  vortex, in comparison, is not flexible
enough for varying $M$ because the symmetric arrangement of the two
singularities in $\phi_1$ and  $\phi_{-1}$ around the center (see Fig.4) becomes 
unbalanced as $\phi_1$ component grows while  $\phi_{-1}$ shrinks.
Thus it is confined in a narrow thin region near $M/N\sim0$ (see Fig.3).

 The  $\langle 1,0,-1\rangle$  vortex, (not shown here,  see Fig.3 in Ref. \cite{isoshima1})
 is stabilized in larger $M/N$ and higher $\Omega$ region because in this vortex
 the dominant component $\phi_1$ has the winding number 1, which can
 effectively lower the rotation energy.

%===================== Figure 4 ======================
\begin{figure}
%\begin{center}
\includegraphics[width=5cm]{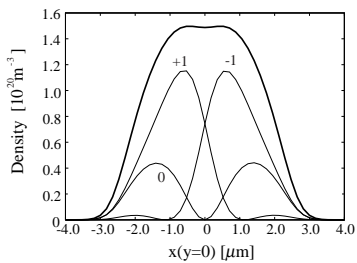}
%\end{center}
\caption{\label{fig:Fdiagram}
Density profile of the condensates 
for the non-axis-symmetric $\langle 1,1,1\rangle$ vortex 
in $\Omega=0.35$ and  M/N=0.0.
The bold line is the total density $\sum_j | \phi_j |^2$
and the thin lines show the density of each component $|\phi_j|^2$.
}
\end{figure}
%===================== Figure 4 ======================/

Having finished the ``global'' stability of the MH and AT vortices, we now turn 
to the ``local'' stability, that is, the stability against a small
perturbation. This is done by solving the extended Bogoliubov equations 
to the three components under the axis-symmetric situation\cite{isoshima1}:
%\begin{eqnarray*}
$
\sum_j \{
        A_{ij}     u_q({\bf r}, j)
      - B_{ij}     v_q({\bf r}, j)\} =\varepsilon_q u_q({\bf r}, i) \\
$, and $
\sum_j \{
        B^{\ast}_{ij}   u_q({\bf r}, j)
      - A^{\ast}_{ij}   v_q({\bf r}, j)\}= \varepsilon_q v_q({\bf r}, i)
$
%\end{eqnarray*}
%
where
$A_{ij} =
      h({\bf r}) \delta_{ij} - \mu_i \delta_{ij}
+ g_n \left\{
         \sum_k  |\phi_k|^2  \delta_{ij}
         + \phi_i \phi_j^{\ast}
      \right\}
      + g_{\rm s}
         \sum_{\alpha}\sum_{kl} [
            (F_{\alpha})_{ij} (F_{\alpha})_{kl} \phi_k^{\ast} \phi_l
            +(F_{\alpha})_{il} (F_{\alpha})_{kj} \left(
               \phi_k^{\ast} \phi_l
            \right)
         ],
$
$
   B_{ij} =
      g_{\rm n} \phi_i \phi_j
      +g_{\rm s} 
         \sum_{\alpha} \sum_{kl} \left[
            (F_{\alpha})_{ik} \phi_k  (F_{\alpha})_{jl} \phi_l
         \right],
$
$u_q({\bf r}, i)$ and $v_q({\bf r}, i)$ are
the $q$-th eigenfunctions with the spin $i$ and
$\varepsilon_q$ corresponds to the $q$-th eigenvalue.
Since this gives the excitation spectrum of the system, the negative energy relative
to the condensation energy at zero implies the local intrinsic instability
of the putative vortex in the energy landscape. We have performed the extensive 
computation to check this local stability for MH vortex and other axis-symmetric
 vortices against a small perturbation. 
 As  expected, the lower phase boundary of the MH in Fig.3 coincides almost completely with the 
 local stability region, below which the lowest excitation mode with
 the angular momentum $q_{\theta}=-1$ for the positive external rotation
 (${\vec \Omega}>0$) becomes negative (see the inset in Fig.3).
 The upper boundary of the $\langle 0, 1, 2 \rangle$ vortex in Fig.3 indicates that
 the lowest excitation mode with mainly $q_{\theta}=+1$
 becomes negative (see the inset in Fig.3). Thus the global stability mentioned roughly corresponds
 to the present local stability. Therefore, it is concluded that the MH and AT vortices
 are stable and robust in the ferromagnetic spinor BEC.
 
 It is easy to calculate the total angular momentum $L$ in MH vortex
 which is given by $L/\hbar=N_0+2N_{-1}$ ($N_i$ is the particle 
 number of the $i$-component). Since $M=N_1-N_{-1}$ and 
 $N=N_1+N_0+N_{-1}$, the total angular momentum per particle is found
 to be ${L\over \hbar N}=1-{M\over N}$. This simple formula means that 
 at $M=0$,  $\phi_0$ with $w_0=1$ carries all the angular momentum 
 as in the usual scalar vortex, and at $M=N$, $L=0$ because 
 $\phi_1$ has no winding. At ${M\over N}={1\over 2}$,  ${L\over \hbar N}$
 is half, implying that in the MH vortex the angular momentum per particle
 is exactly $\hbar/2$.
 
 In summary, we have shown that the Mermin-Ho and Anderson-Toulouse
 vortices are thermodynamically stable under a certain rotation drive and
 the total magnetization of the system, both of which are experimentally well-controlled 
 parameters. We have examined their stability by two methods, local and 
 global ones in the energy landscape, coinciding with the earlier conclusion
 that these are unstable under no rotation drive\cite{stoof}. These intriguing objects might be
 detected by various ways, such as optically by utilizing the Faraday rotation
 which images the spatial magnetic pattern of these vortices.  
 A favorable experiment is to use an oblate ellipsoidal shaped system
 to mimic our disk calculation. An elongated cigar system may not be appropriate, which
 induces the phase separation along the long axis.

 We thank T. Ohmi and T. Isoshima for valuable discussions.

%%%% references %%%%%%%%%%%%%%%%%%%%%%%%%%%%%%%
%%%%%%%%
%\begin{references}

%\end{references}
%%%%%%%%%%%%%%

%Fig.1:
%Density profile of the condensates (a) 
%and the density map of $l_z$-vector (b) 
%for the $\langle 0,1,2\rangle$ vortex at $\Omega=0.35$
%and M/N=0.21.
%The bold line is the total density $\sum_j | \phi_j |^2$ 
%and the thin lines show the density of each component $|\phi_j|^2$.

%Fig.2:
%Spatial dependence of the $l_z$-component along the radial direction 
%at M/N=0, 0.46.
%The dashed line shows cos$\beta(r)$ with the bending angle
%$\beta(r)=\pi r/R$ ($R=3.0\mu m$)
%for the AT vortex.

%Fig.3:
%Phase diagram in the ferromagnetic state.
%The lowest eigenvalues along $q_{\theta}$ at $\Omega$=0.3 
%are displayed in the inset. 
%Triangles: M/N=0.2 and filled triangles: 0.75.

%Fig.4:
%Density profile of the condensates 
%for the $\langle 1,1,1\rangle$ vortex in $\Omega=0.35$ and  M/N=0.0.
%The bold line is the total density $\sum_j | \phi_j |^2$
%and the thin lines show the density of each component $|\phi_j|^2$.

\end{document}